\documentclass[a4paper,11pt]{article}
\usepackage{jcappub} % for details on the use of the package, please
\usepackage[T1]{fontenc} % if needed

\usepackage{fancyhdr}
\usepackage{amssymb}
\usepackage{amsmath}
\usepackage{tikz}
\usepackage{graphicx}
\usepackage{multirow}

\newcommand{\eq}[1]{Eq.~(\ref{#1})}
\newcommand{\eqs}[2]{Eqs.~(\ref{#1}) and (\ref{#2})}
\newcommand{\eqss}[3]{Eqs.~(\ref{#1}), (\ref{#2}) and (\ref{#3})}

\newcommand{\GeV}{\mathinner{\mathrm{GeV}}}

\def\bea{\begin{eqnarray}}
\def\eea{\end{eqnarray}}
\def\beq{\begin{equation}}
\def\eeq{\end{equation}}

\def\l{\left}
\def\r{\right}

\subheader{\small \hfill IFIC/19-11}

\title{\boldmath Spontaneous baryogenesis in \textit{spiral inflation}}

\author[1]{Gabriela Barenboim}
\author[2]{and Wan-Il Park}
\affiliation[1]{Departament de F\'isica Te\`orica and IFIC, Universitat de Val\`encia-CSIC, \\ E-46100, Burjassot, Spain}
\affiliation[2]{Division of Science Education and Institute of Fusion Science, Chonbuk National University, \\ Jeonju 54896, Korea}

\emailAdd{Gabriela.Barenboim@uv.es}
\emailAdd{wipark@jbnu.ac.kr}

\abstract{We examined the possibility of spontaneous baryogenesis driven by the inflaton in the scenario of \textit{spiral inflation}, and found the parametric dependence of the 
late-time baryon number asymmetry.
As a result, it is shown that, depending on the effective coupling of baryon/lepton number violating operators, it is possible to obtain the right amount of asymmetry even in the presence of a matter-domination era as long as such era is relatively short.
In a part of the parameter space, the required expansion rate during inflation is close to the current upper-bound, and hence can be probed in the near future experiments.
}

\begin{document}
\maketitle
\flushbottom

\section{Introduction}
It is known that our visible world is made of either matter or anti-matter only, depending on how we define them.
Observations indicate that the asymmetry between matter and anti-matter in terms of the ratio of baryon-to-entropy density is about $10^{-10}$ \cite{pdg-bbn,Aghanim:2018eyx}.
Such an asymmetry could be an initial condition of the universe evolving to our present universe.
However, in the presence of inflation \cite{Guth:1980zm,Sato:1980yn,Starobinsky:1980te} which is now believed to be a crucial ingredient of the thermal  history of the universe at a very early epoch, typically well before the conventional electroweak phase transition, an initial asymmetry which might have existed is expected to be diluted to a totally negligible level, and hence there should be a process, called baryogenesis, able to generate an asymmetry after inflation.

Typically, when it works through particle-interactions, a baryo/leptogenesis mechanism is required to satisfy the so called Sakharov conditions \cite{Sakharov:1967dj}, i.e., (i) baryon($B$)/lepton($L$) number violating process, (ii) $C$- and $CP$-violation, and (iii) out-of-equilibrium decay of particles producing baryon/lepton number.
However, when the dynamics of a background field is involved, the above conditions can be relaxed.
Spontaneous baryogenesis \cite{Cohen:1987vi,Cohen:1988kt} (see also Ref.~\cite{DeSimone:2016ofp} for cosmological aspects), is a specially interesting case  as the asymetry can be generated in equilibrium.
The key feature of spontaneous baryogenesis is that, when a baryon/lepton current is coupled to a background evolution of a field, the time-dependence of the field  can provide an effective chemical potential associated with baryon/lepton number.
As a result, in the presence of $B$- or $L$-violating processes in thermal equilibrium, an asymmetry of $B$- or $L$-number can be generated even in the thermal bath. 
The main question in this novel scenario is the identity of the background field and its precise nature which should allow $B$- or $L$-violating processes in thermal equilibrium.
In principle, the background field can be any scalar field which has a sizable time-evolution at the epoch of baryo/leptogenesis as long as the symmetries of the theory allow a time-dependent coupling of the field to the baryonic/leptonic current.

An additional possibility related to spontaneous baryogenesis is the production of an asymmetry from the decay of the oscillating scalar field associated with spontaneous baryogenesis.
Typically, the nature of the deriving field for spontaneous baryogenesis is an angular degree of freedom of a complex field.
Hence, when the complex field carries a charge, a motion of the phase field implies an asymmetry of the charge.
Even though an oscillation of the angular degree with respect to a true vacuum can not provide an asymmetry with a definite sign, its decay can results in a net baryon/lepton asymmetry with a specific sign, thanks to the expansion of the universe \cite{Cohen:1987vi} \footnote{See for example Refs.~\cite{Anber:2015yca,Cado:2016kdp,Takahashi:2015ula} for other possibilities utilizing pseudo-scalar inflaton}. 
A model-dependent question is if the net asymmetry can be large enough to match observations.   

On the other hand, \textit{Spiral inflation} \cite{Barenboim:2014vea,Barenboim:2015zka} was proposed as a phenomenological scenario of inflation circumventing the flatness and trans-Planckian issues of the inflaton potential.
One of the key features of such a  scenario is that the inflaton trajectory is spiraling-out and inflation ends by a waterfall-like drop.
Such a spiral motion is something similar to an angular motion of a complex field.
Hence, a natural question  is whether the inflaton in the  \textit{Spiral inflation} scenario can be responsible for generating the right amount of baryon number asymmetry through either spontaneous baryogenesis or the second possibility mentioned in the previous paragraph, {\it i.e.} through its decays in an expanding Universe.
%
%is associated with a complex field, it may have a derivative coupling to matter current.
%As a result, inflaton can be used for spontaneous baryogenesis.

In this work, within the  framework of spiral inflation, we show that a right amount of baryon number asymmetry can be achieved in a certain parameter space not by spontaneous baryogenesis but by the remnant of the decays of the inflaton.
This paper is organized as follows.
In section~\ref{sec:model}, the general form for the potential for spiral inflation is introduced.
In section~\ref{sec:spiral-inflation}, spiral inflation is described, including the post-inflation behavior of the field configuration.
In section~\ref{sec:asymmetry}, the genesis of charge asymmetry by means of the inflaton after inflation through either spontaneous baryogenesis or decays of inflaton is considered, searching for the parameter space for a right amount of baryon number asymmetry at the present universe.  
In section~\ref{sec:con}, conclusions are drawn.

\section{The model}
\label{sec:model}

The general form of the potential responsible for \textit{spiral infaltion} can be written as 
\beq \label{eq:V}
V = V_\phi + V_{\rm m}
\eeq
where
\bea
V_\phi &=& V_0 \l[ 1 - f(\phi) \r]^2
\\
V_{\rm m} &=& \Lambda^4 \l[ 1 - \cos \l( h(\phi) - \theta \r) \r]
\eea
with
\beq
f(\phi) = \l( \phi/\phi_0 \r)^p, \quad h(\phi) = \l( \phi/M \r)^q
\eeq
and $p,q>0$.
We may take $p$ and $q$ to be non-negative integers, and consider the case of $p\geq4$ and $0<q\leq2$ which might be theoretically plausible. 
Clearly $\phi$ and $\theta$ can be considered as the modulus and the phase of a complex field given by $\Phi = \phi e^{i \theta} / \sqrt{2}$.
This potential is a hilltop potential having a trench \textit{spiraling-out} from the hilltop.

Starting from the hilltop, the field configuration is expected to follow closely the minimum of the trench as long as the curvature along the orthogonal direction is large enough, satisfying $V'=0$ with `$\prime$' denoting a derivative with respect to $\phi$.
Hence, from 
\bea
V' &=& V_\phi' + V_{\rm m}' = - 2 V_0 \l( 1 - f \r) f' + \Lambda^4 h' \sin \l( h - \theta \r)
\\
V'' &=& V_\phi'' + V_{\rm m}'' = - 2 V_0 \l[ \l( 1 - f \r) f'' - \l( f' \r)^2 \r] + \Lambda^4 \l[ \l( h' \r)^2 \cos \l( h - \theta \r) + h'' \sin \l( h - \theta \r) \r]
\\
\frac{\partial V'}{\phi \partial \theta} &=& \frac{\partial V_{\rm m}'}{\phi \partial \theta} = - \frac{\Lambda^4}{\phi} h' \cos \l( h - \theta \r)
\\
\frac{\partial^2 V}{\phi^2 \partial \theta^2} &=& \frac{\partial^2 V_{\rm m}}{\phi^2 \partial \theta^2} = \frac{\Lambda^4}{\phi^2} \cos \l( h - \theta \r) 
\eea
one finds that along the trajectory
\beq \label{eq:min-cond}
V_\phi' = - V_{\rm m}' \quad \Leftrightarrow \quad  2 V_0 \l( 1 - f \r) f' = \Lambda^4 h' \sin \l( h - \theta \r)
\eeq
leading to 
\beq
V'' d\phi = - \l( \partial V_{\rm m}' / \partial \theta \r) d \theta
\eeq
Therefore, for $|V_\phi''| \ll |V_{\rm m}''|$ which is expected to be satisfied in the vicinity of the minimum of the trench, 
\beq
\frac{d\phi}{d \theta} \simeq \frac{h'}{\l( h' \r)^2 + h'' \tan \l( h - \theta \r)}
\eeq

We denote the trajectory following the minimum of trench and the direction orthogonal to the trajectory as $I$ and $\psi$, respectively.
Then, an infinitesimal displacement along $I$ can be written as
\beq
dI \equiv \l[ 1 + \l( \frac{\phi d \theta}{d \phi} \r)^2 \r]^{1/2} d\phi = \l[ 1 + \l( \frac{d\phi}{\phi d\theta} \r)^2 \r]^{1/2} d\theta
\eeq
with the unit vectors along $I$ and $\psi$ given by,
\beq
\bold{e}_I^T = \l( c_\phi, c_\theta \r), \quad \bold{e}_\psi^T = \l( c_\theta, -c_\phi \r)
\eeq
so that the directional derivatives are found to be respectivcely
\bea
\frac{d}{dI} &=& \bold{e}_I \cdot \nabla = c_\phi \frac{\partial}{\partial \phi} + c_\theta \frac{\partial}{\partial \theta}
\\
\frac{d}{d\psi} &=& \bold{e}_\psi \cdot \nabla = c_\theta \frac{\partial}{\partial \phi} - c_\phi \frac{\partial}{\partial \theta}
\eea
where
\bea
c_\phi &\equiv& \frac{\partial \phi}{\partial I} = \frac{d \phi/d \theta}{\sqrt{\phi^2 + \l( d \phi/d \theta \r)^2}}
\\
c_\theta &\equiv& \frac{\phi \partial \theta}{\partial I} = \frac{\phi}{\sqrt{\phi^2 + \l( d \phi/d \theta \r)^2}} 
\eea
Hence, one finds
\bea
\frac{dV}{dI} &=& c_\phi \frac{\partial V}{\partial \phi} + c_\theta \frac{\partial V}{\phi d \theta} 
\\
\frac{dV}{d\psi} &=& c_\theta \frac{\partial V}{\partial \phi} - c_\phi \frac{\partial V}{\phi d \theta} 
\eea
and
\bea
\frac{d^2 V}{d I^2} &=& c_\phi^2 \mathbb{M}_{\phi \phi}^2 + 2 c_\phi c_\theta \mathbb{M}_{\phi \theta}^2 + c_\theta^2 \mathbb{M}_{\theta \theta}^2
\\
\frac{d^2 V}{d \psi^2} &=& c_\theta^2 \mathbb{M}_{\phi \phi}^2 - 2 c_\phi c_\theta \mathbb{M}_{\phi \theta}^2 + c_\phi^2 \mathbb{M}_{\theta \theta}^2
\eea
where the elements of the mass-square matrix ($\mathbb{M}^2$) are found to be
\bea
\mathbb{M}^2_{\phi \phi} &=& \frac{\partial^2 V}{\partial \phi^2} + \frac{\partial \ln c_\phi}{\partial \phi} \frac{\partial V}{\partial \phi} 
\\
\mathbb{M}^2_{\phi \theta} &=& \frac{\partial^2 V}{\phi \partial \theta \partial \phi} + \frac{1}{2} \l( \frac{\partial \ln c_\theta}{\partial \ln \phi} -1 \r) \frac{\partial V}{\phi^2 \partial \theta} + \frac{1}{2} \frac{\partial \ln c_\phi}{\partial \theta} \frac{\partial V}{\phi \partial \phi}
\\
\mathbb{M}^2_{\theta \theta} &=& \frac{\partial^2 V}{\phi^2 \partial \theta^2} + \frac{\partial \ln c_\theta}{\partial \theta} \frac{\partial V}{\phi^2 \partial \theta}
\eea

%Starting from around the hilltop, the field configuration spirals out, following closely the minimum of trench as long as the curvature along the orthogonal direction $\psi$ is large enough \cite{}.
%Spiral inflation takes place during this spiral motion.
%Inflation ends as the field configuration exits trench, falling along $\phi$ direction at a point satisfying 

Spiral motion ends as the field configuration leaves the trench, falling along the  $\phi$ direction at a point satisfying 
\beq \label{eq:phie-cond}
2 V_0 \l( 1 - f \r) f' = \Lambda^4 h'
\eeq
There are two solutions of \eq{eq:phie-cond}, denoted as $\phi_e$ and $\phi_{\rm r}$, % with $\phi_e < \phi_{\rm r}$. 
The smaller one is $\phi_e$, the end point of the slow-roll inflation.
For $p \geq 4$ which is the case we are interested in, $f_e \ll 1$ unless $\phi_e$ is quite close to $\phi_0$.
In this case, 
\beq \label{eq:phie}
\frac{\phi_e}{\phi_0} = \l[ \frac{qh_0}{2p \l( 1 - f_e \r)} \frac{\Lambda^4}{V_0} \r]^{\frac{1}{p-q}} 
\simeq \l[ \frac{qh_0}{2p} \frac{\Lambda^4}{V_0} \r]^{\frac{1}{p-q}} 
\equiv \kappa^{\frac{1}{p-q}}
\eeq
where $h_0 \equiv (\phi_0/M)^q$ and 
\beq
\kappa \equiv \frac{qh_0}{2p} \frac{\Lambda^4}{V_0}
\eeq
which is determined by $\Lambda$ for a given choice of the other parameters.
Note that, as long as $\kappa \ll 1$, the approximation in \eq{eq:phie} is good enough for our purpose.

The other solution, the largest one, $\phi_{\rm r}$, represents the location of the re-trapping of the field in the trench.
It satisfies
\beq \label{eq:phi-retrap}
\frac{\phi_{\rm r}}{\phi_0} = \l[ 1- \frac{\Lambda^4}{2V_0} \frac{h_{\rm r}'}{f_{\rm r}'} \r]^{1/p} = \l[ 1- \kappa \l( \frac{\phi_0}{\phi_{\rm r}} \r)^{p-q} \r]^{1/p} \simeq 1 - \frac{\kappa}{p}
\eeq
where again we assumed $\kappa \ll 1 \ll p$, resulting in $\phi_{\rm r} \approx \phi_0$.

%In case of retrapping, there can be a second stage of inflation adding extra $e$-foldings although it depends on the size of the second term in \eq{eq:phi-retrap}.
%In the subsequent discussion, we consider the case of 
%\beq
%\kappa \lesssim \mathcal{O}(10^{-3})
%\eeq
%in order to avoid complication of inflationary dynamics caused by early re-trapping.

\section{Spiral inflation}
\label{sec:spiral-inflation}

\subsection{Inflation ($\phi<\phi_e$)}
When $\phi < \phi_e$,  inflation consistent with observations  (in slow-roll regime) takes place due to the gentle  spiral dynamics of the field configuration. 
For $q \leq 2$ and $M \ll \phi < \phi_e$, if $\phi$ is away enough from $\phi_e$, it is expected that $f(\phi) \ll f_e \ll 1$ leading to $|V_\phi'(\phi)| \ll |V_\phi'(\phi_e)|$.
In this case, we expect $\sin \l( h - \theta \r) \lesssim \cos \l( h - \theta \r)$.
Hence 
\beq \label{eq:dphi-dth}
\frac{d\phi}{d\theta} \simeq \frac{h'}{ \l( h' \r)^2 + h'' \tan \l( h - \theta \r)} \simeq \frac{1}{h'}
\eeq
and 
\beq \label{eq:cphi-cth}
c_\phi \simeq \frac{1}{q h}, \quad c_\theta \simeq \l( 1 + \frac{1}{(qh)^2} \r)^{-1/2} \simeq 1
\eeq
where $qh \gg 1$ was used.
Hence, using \eq{eq:min-cond}, we find
\bea
\frac{dV}{dI} 
&\simeq& - \frac{2p (1-f) f}{(qh)} \frac{V_0}{\phi} 
\\
\frac{d^2 V}{d I^2}  \label{eq:mIsq}
&\simeq& - \frac{2V_0}{(qh)^2 \phi^2} \l\{ p \l[ p-(q+1) \r](1-f)f - (pf)^2 \r\}
%\nonumber \\
%&\simeq& -\frac{2p \l[ p-(q+1) \r]f}{\l( qh \r)^2} \frac{V_0}{\phi^2}
\\
\frac{d^2 V}{d \psi^2} &\simeq& (qh)^2 \frac{\Lambda^4}{\phi^2}
\eea
%where in the second line of \eq{eq:mIsq} we used $f(\phi) \ll 1$.
Note that in order for $I$ to follow closely the minimum of the trench during inflation, the mass scale along $\psi$ should be large enough or at least comparable to the expansion rate, that is, $m_{\psi}^2/3 H_*^2 \gtrsim 1$ which constrains $\kappa$ to satisfy
\beq \label{eq:kappa-bnd}
\kappa \gtrsim \frac{1}{2pq h_0} \l( \frac{\phi_0}{M_{\rm P}} \r)^2 \l( \frac{\phi_0}{\phi_*} \r)^{2(q-1)}
\eeq
where `$_*$' denotes a quantity associated with a pivot scale of observations.
It turns out that this constraint is easily satisfied in the parameter space we are interested in. 
The slow-roll parameters are given by
\bea \label{eq:eta}
\eta &\simeq& - \frac{2p \l[ p - (q+1) \r]}{(qh_0)^2} \l( \frac{M_{\rm P}}{\phi_0} \r)^2 \l( \frac{\phi}{\phi_0} \r)^{p-2(q+1)} \equiv - g(M, \phi_0) \l( \frac{\phi}{\phi_0} \r)^{p-2(q+1)}
\\ \label{eq:epsilon}
\epsilon &\simeq& - \frac{pf}{p - (q+1)} \eta
\eea
where we used $f(\phi) \ll 1$ and defined a function $g(M, \phi_0)$ which will prove to be convenient later.

At $\phi = \phi_*$ associated with, for example, Planck pivot scale, $f(\phi_*) \ll 1$ is expected (see \eq{eq:phie}).
Hence, $\epsilon_* \ll |\eta_*|$ as long as $p/\l[p-(q+1) \r] \sim \mathcal{O}(1)$, and $\eta_*$ is nearly fixed by the observed spectral index of the density power-spectrum as $n_s^{\rm obs} \simeq 1 + 2 \eta_*$ in order to match observations.
The power spectrum is given by 
\beq
P_R = \frac{H^2}{8 \pi^2 \epsilon M_{\rm P}^2}
\eeq
leading to
\beq \label{eq:Hstar}
\frac{H_*}{M_{\rm p}} = \l[ \frac{8 \pi^2 P_R^{\rm obs} p |\eta_*|}{p-(q+1)} \r]^{1/2} \l( \frac{\phi_*}{\phi_0} \r)^{p/2}
\eeq 
where $P_R^{\rm obs} = 2.1\times 10^{-9}$ \cite{Aghanim:2018eyx} is the observed amplitude of the  density power spectrum.
If \eq{eq:dphi-dth} is satisfied for most of the region of $(\phi_*, \phi_e)$, the number of $e$-foldings generated is found to be
\beq \label{eq:Ne}
N_e \simeq \l\{
\begin{array}{ll}
\frac{p-(q+1)}{|\eta_*|} \ln \l( \frac{\phi_e}{\phi_*} \r) & {\rm for} \  p = 2(q+1)
\\
\frac{p-(q+1)}{p-2(q+1)} \frac{1}{|\eta_*|} \l[ 1 - \l( \frac{\phi_*}{\phi_e} \r)^{p-2(q+1)} \r] & {\rm for} \ p \neq 2(q+1)
\end{array}
\r.
\eeq
For a given comoving scale $k_*$ and the present horizon $k_0$, observations require such number of $e$-foldings to be
\beq \label{eq:Ne-obs}
N_e^{\rm obs}(k_*) 
= 62 + \ln \frac{k_0}{k_*} - \ln \l( \frac{10^{16} \GeV}{V_*^{1/4}} \r) - \frac{1}{3} \ln \l( \frac{V_*^{1/4}}{\rho_{\rm R}^{1/4}} \r)
\eeq
where we took $V_e = V_*$ and $\rho_{\rm R}$ is the radiation energy density when reheating is efficient enough to recover a radiation-dominant universe.
Specifically, we take $\rho_{\rm R}$ to be the energy density of the universe when $H = (2/3) \Gamma_\psi$ with $\Gamma_\psi$ being the decay rate of $\psi$-particles.

\begin{figure}[h]
\begin{center}
\includegraphics[width=0.45\textwidth]{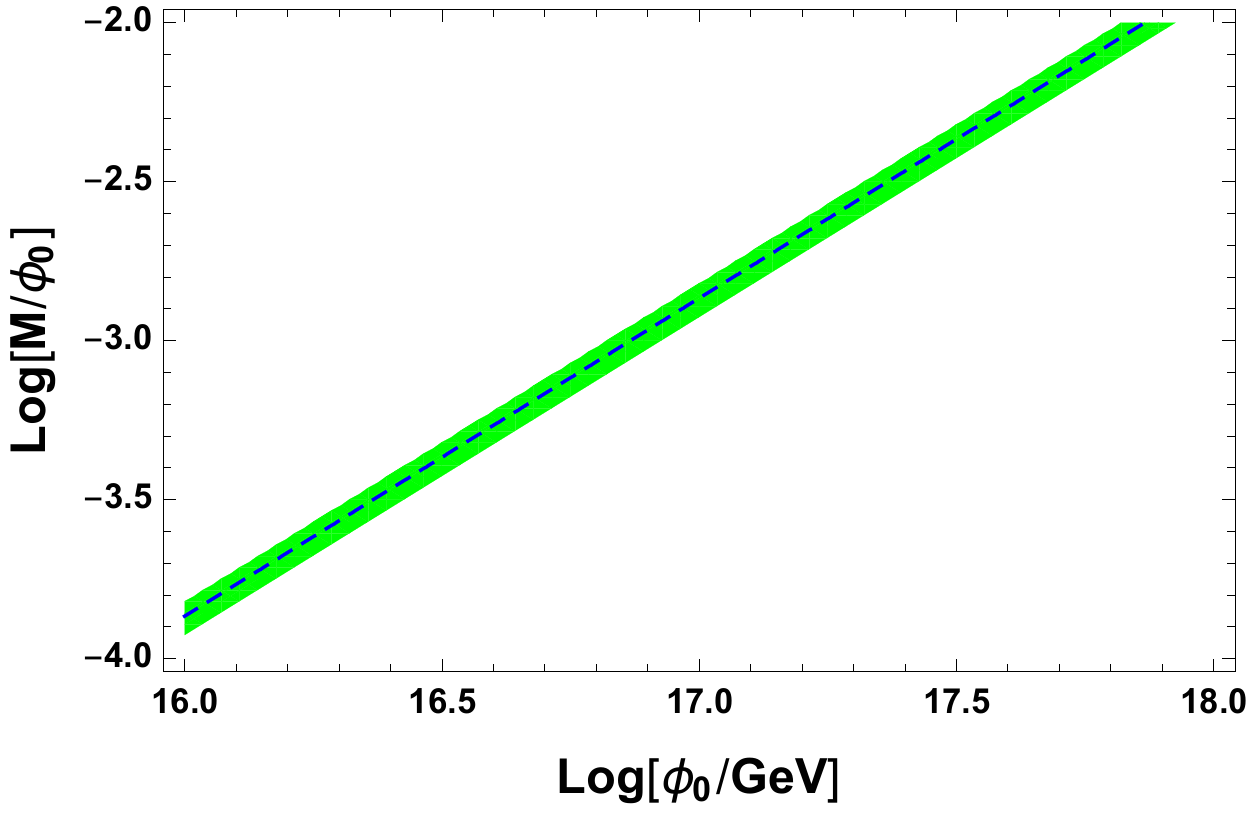}
\includegraphics[width=0.45\textwidth]{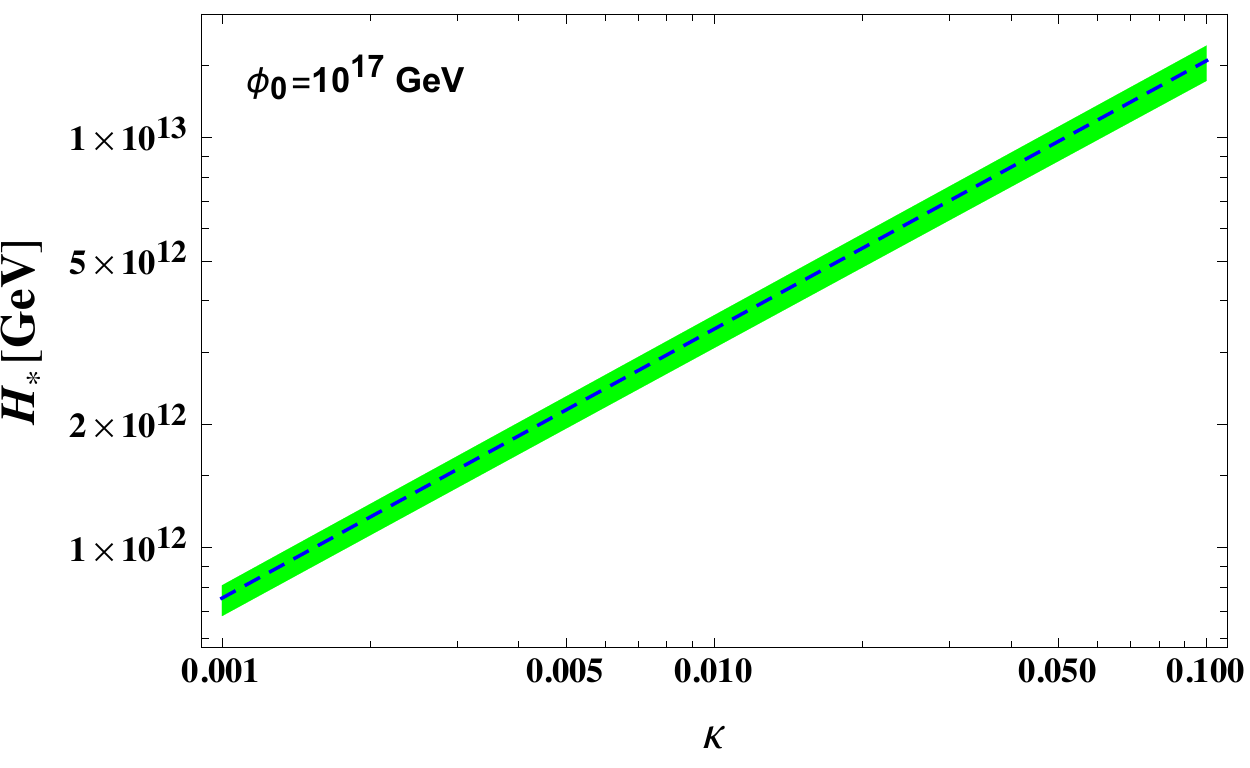}
\includegraphics[width=0.45\textwidth]{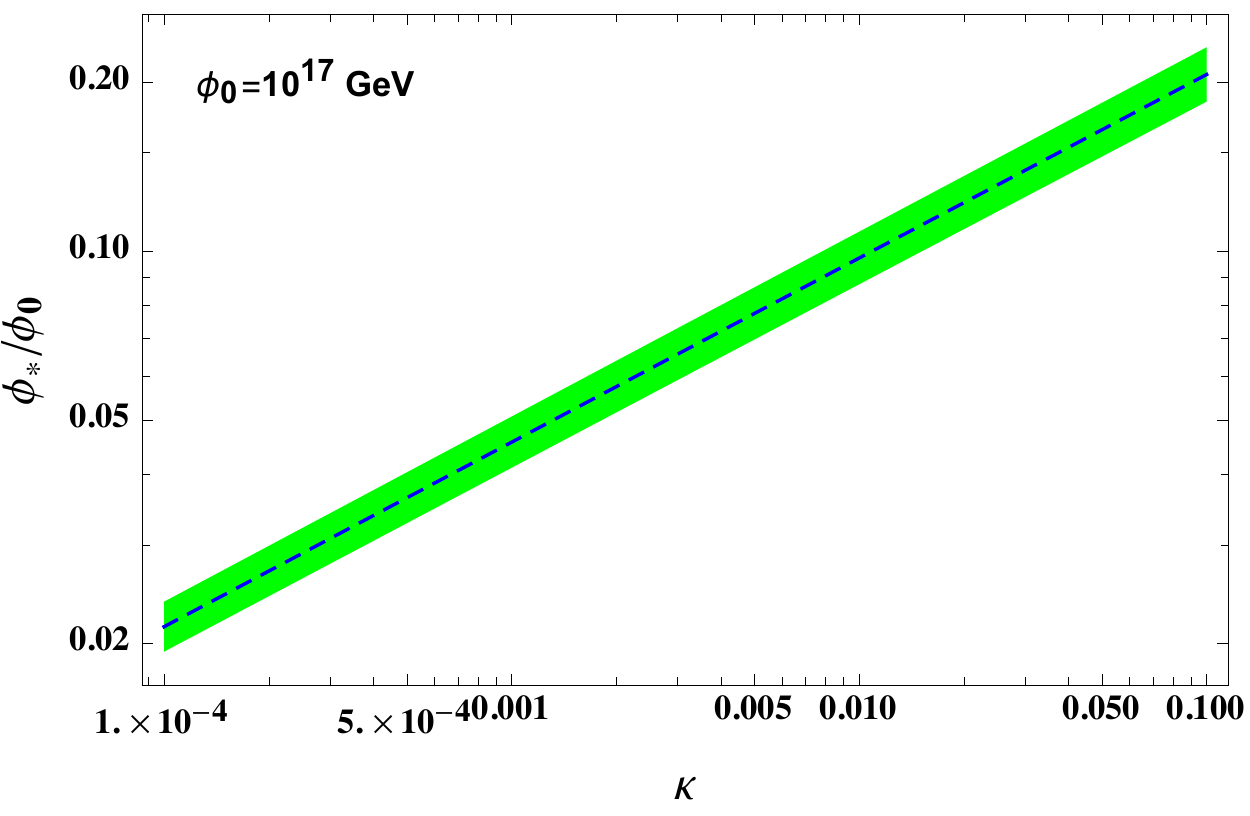}
\includegraphics[width=0.45\textwidth]{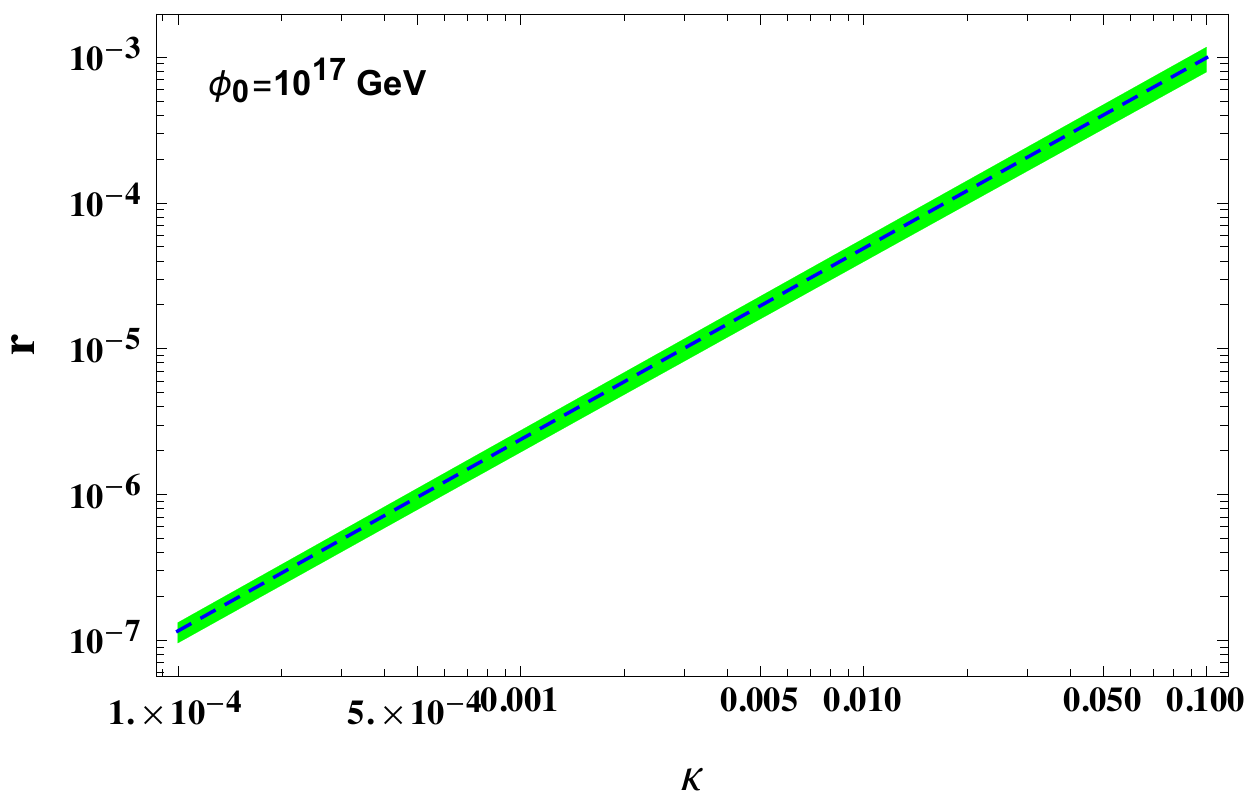}
\caption{Parameter spaces matching observations for the case $p=2(q+1)=4$.
\textit{Top-Left}: $(\phi_0,M)$ satisfying $g(M, \phi_0)=|\eta_*|$,
\textit{Top-Right}: $(\kappa,H_*)$,
\textit{Bottom-Left}: $(\kappa, \phi_*/\phi_0)$,
\textit{Bottom-Right}: $(\kappa, r)$.
In each panel, the dashed line corresponds to the case of $n_s = 0.9659$, and the green band represents the $2$-$\sigma$ allowed region of $n_s$.}
\label{fig:low-p-parameters}
\end{center}
\end{figure}

For a given set of $(p,q)$, if the model-dependent couplings of $\phi$ and $\theta$ to other matter fields are fixed, the model parameters which still remain free are :
\beq
V_0, \Lambda, \phi_0, M
\eeq
Also, there are three observable constraints:
\beq
n_s^{\rm obs}, P_\mathcal{R}^{\rm obs}, N_e^{\rm obs}
\eeq
%where $n_s^{\rm obs}$ is the observed spectral index of the density power-spectrum. 
Those free parameters and observables are related by four equations, \eq{eq:phie}, (\ref{eq:eta}), (\ref{eq:Hstar}), and \eq{eq:Ne} equated with \eq{eq:Ne-obs}.
Hence, all the free parameters are fixed by the  observables as follows.
First of all, as shown in \eq{eq:epsilon}, $\epsilon_* \ll \eta_*$ which leads to $n_s^{\rm obs} \simeq 1 + 2 \eta_*$.
If $p=2(q+1)$, from \eq{eq:eta} $g(M, \phi_0)$ is fixed, determining $M$ as a function of $\phi_0$.
Plugging \eq{eq:Hstar} into \eqs{eq:Ne}{eq:Ne-obs} with $\Gamma_\psi$ being expressed as a function of $M$ and $\phi_0$ as shown in the next subsection, $H_*$ is determined as
\bea \label{eq:HIL}
\ln \frac{H_*}{M_{\rm P}} 
&=& \l( \frac{1}{|\eta_*|} + \frac{5}{6} \r)^{-1} \l\{ \frac{p}{2 |\eta_*|} \l[ \frac{\ln \kappa}{p-q} + \frac{\ln \l( 16 \pi^2 P_R^{\rm obs} |\eta_*| \r)}{p} \r] \r.
\nonumber \\
&& \l. - \l[ 62 + \ln \frac{k_0}{k_*} - \ln \l( \frac{10^{16} \GeV}{3^{1/4} M_{\rm P}} \r) + \frac{1}{6} \ln \l( \frac{\gamma_\psi}{8 \pi} \l( 6 p^2 \l(1+\frac{\kappa qh_0}{p} \r) \r)^{3/2} \l( \frac{M_{\rm P}}{\phi_0} \r)^5 \r) \r] \r\}
\eea
which can be re-used to find $\phi_*$, using \eq{eq:Hstar}.
Note that $\kappa$ is treated as a free parameter in this case although it should satisfy \eq{eq:kappa-bnd}. 
Also, $H_*$ depends dominantly on $\kappa$ due to the factor $1/|\eta_*|$ in front of $\ln \kappa$ in \eq{eq:HIL}. 
In Fig.~\ref{fig:low-p-parameters}, we show the allowed parameter spaces.
As can be clearly seen in the figure, the $\phi_0$ dependence of each parameter except $M$ is quite weak, so, for simplicity we took $\phi_0 = 10^{17} \GeV$ as a representative value.
\begin{figure}[h]
\begin{center}
\includegraphics[width=0.45\textwidth]{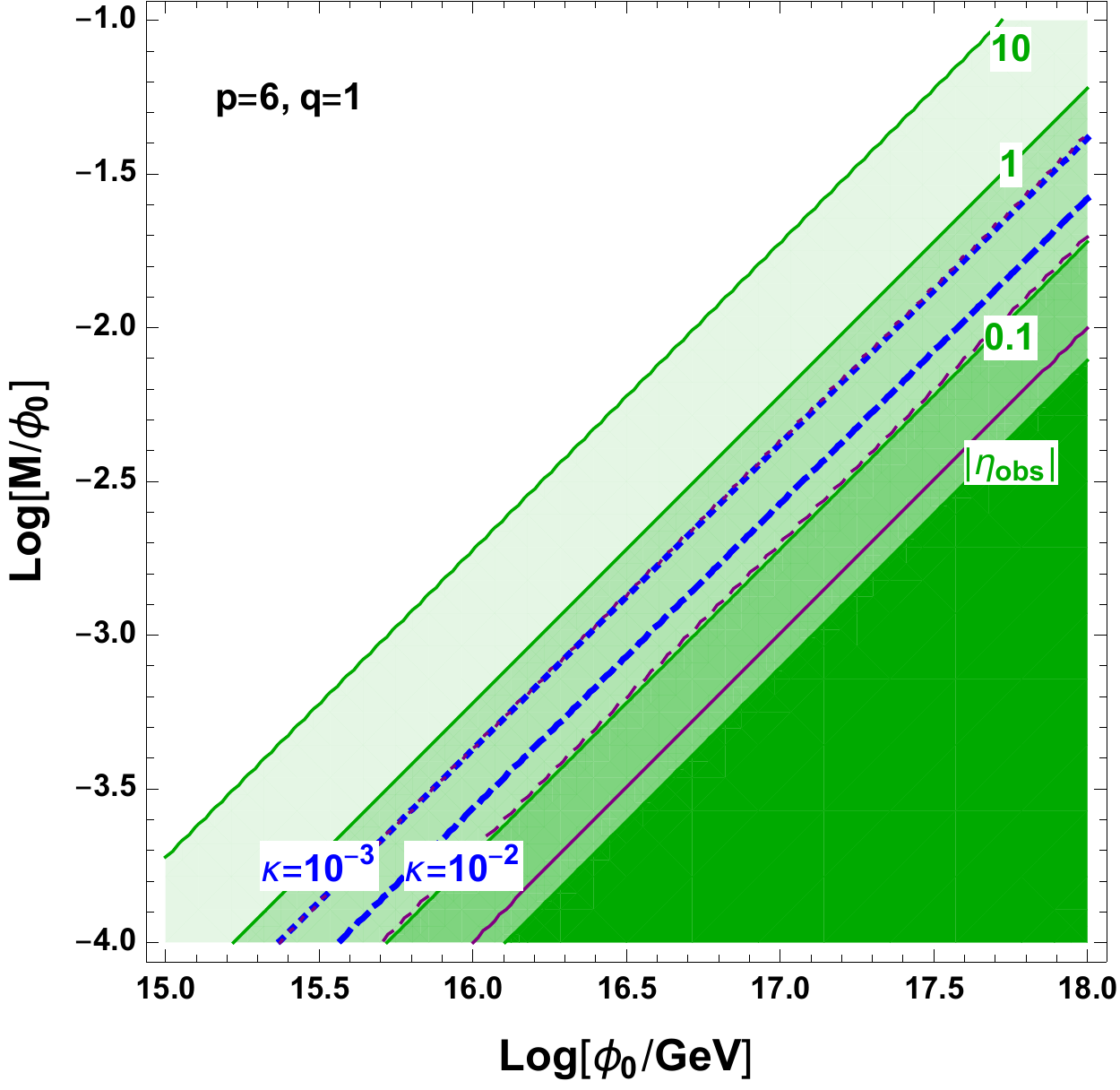}
\includegraphics[width=0.45\textwidth]{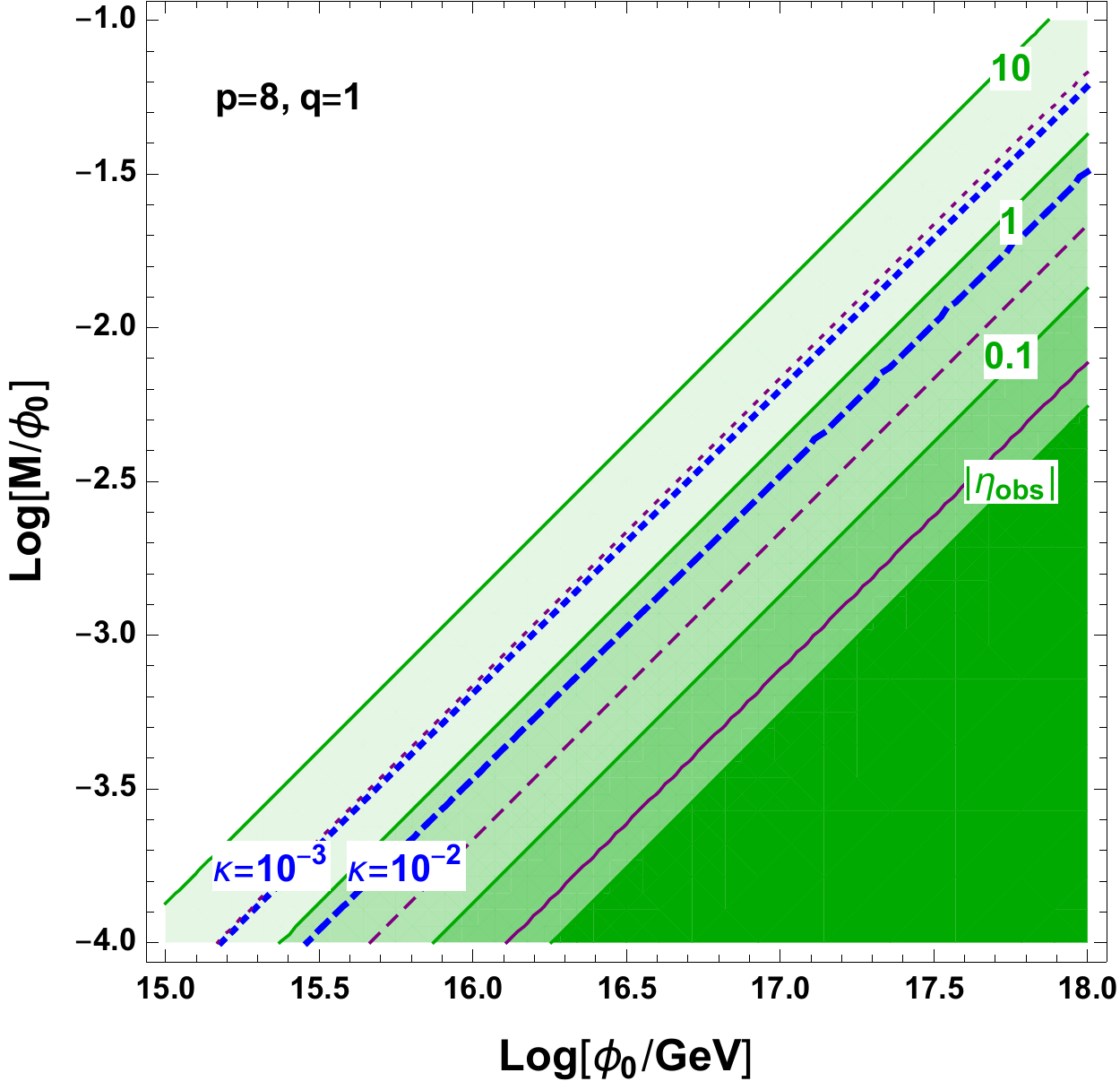}
\caption{Parameter spaces (blue dashed and dotted lines corresponding to $\kappa = 10^{-2}, 10^{-3}$, respectively) matching observations for the cases $(p,q)=(6,1)$ (left) and $(p,q)=(8,1)$ (right).
In both panels, for the decay rate of $\psi$ we used $\gamma_\psi=1$ (see \eq{eq:decay-rates}). 
Green regions cover ranges of $g(M, \phi_0)$ indicated by the number on the green diagonal lines bounding the regions with different transparency leves.
The opaque green region is for $g(M, \phi_0) < |\eta_{\rm obs}|$ and excluded.
Purple solid/dashed/dotted line is for $H_*=H_*^{\rm bnd}/10^{13} \GeV/10^{12} \GeV$. 
As inflationary observabels, $n_s = 0.9659$ and $P_R = 2.1 \times 10^{-9}$ were used \cite{Aghanim:2018eyx}.
2-$\sigma$ uncertainties in inflationary observables do not alter the figure significantly.
The tensor-to-scalar ratios on blue lines are $r \approx 2.5 \times 10^{-4}$ (dashed) and $2 \times 10^{-5}$ (dotted).
}
\label{fig:large-p-parameters}
\end{center}
\end{figure}

If $p > 2(q+1)$, from \eq{eq:eta} one can regard $\phi_*/\phi_0$ as a function of $M$ and $\phi_0$.
For a given set of $(M, \phi_0)$, \eq{eq:Hstar} constrains $H_*$, and hence $N_e^{\rm obs}$ becomes a function of those free parameters.
Then, from \eqs{eq:Ne}{eq:Ne-obs}, $\phi_e/\phi_0$ is constrained for each pair $(M,\phi_0)$.
Therefore, $\kappa$ in \eq{eq:phie} is not a free parameter but should satisfy the following equation,
\beq \label{eq:kappa}
\kappa = \l\{ \frac{|\eta_*|}{g(M,\phi_0) \l[ 1 - \frac{p-2(q+1)}{p-(q+1)} |\eta_*| N_e^{\rm obs}(M, \phi_0,\kappa) \r]}\r\}^{\frac{p-q}{p-2(q+1)}}
\eeq
The $\kappa$-dependece of $N_e^{\rm obs}$ in \eq{eq:kappa} is from the dependence of $\rho_R$ on $\Gamma_\psi$ (see \eqs{eq:mpsi0}{eq:decay-rates})
Since, for a couple of orders of magnitude variation of $\kappa$, the change of $N_e^{\rm obs}$ is of $\mathcal{O}(1)$, we ignore such a dependence for simplicity.
Such an assumption  is equivalent to setting $\kappa=0$ in \eq{eq:mpsi0} when $N_e^{\rm obs}$ is estimated.
\eq{eq:kappa} can be satisfied by adjusting the ratio $\Lambda^4/V_0$ as a free parameter replacing $\Lambda$.
In Fig.~\ref{fig:large-p-parameters}, the parameter space matching inflationary observables is depicted for $(p,q)=(8,1)$ with $\kappa = 10^{-2}$ and $10^{-3}$ as an example.

\subsection{Post inflation ($\phi_e < \phi \sim \phi_0$)}

As the inflaton leaves the  trench at $\phi_e$, the dominant dynamics turns to the oscillation along $\phi$ with respect to $\phi_0$ as long as the oscillation amplitude $\delta \phi$ satisfies $\delta \phi \gg |\phi_0 - \phi_{\rm r} | = (\kappa/p) \phi_0$.
At this period, $V_\phi$ governs the dynamics, i.e., the time scale of the oscillation is determined by the mass scale along $\phi$ at $\phi \approx \phi_0$,
\beq \label{eq:mphi-0}
m_\phi^2 \equiv \l. V_\phi'' \r|_{\phi \approx \phi_0} = \frac{2(pf)^2V_0}{\phi^2}  \l[ 1-\l(1-\frac{q+1}{p} \r) \l( \frac{1}{f} - 1 \r) \r]
%\l[ p^2 \l(2f^2-f \r) - p(q+1) \l( f^2 - f \r) \r]
\eeq 
As the motion along $\phi$ becomes sufficiently small, the field configuration can be re-trapped in the trench and the dynamics would be again along the minimum of the canal in the vicinity of the true vacuum.
In this case, $V_\phi' = - V_{\rm m}'  \simeq 0$, and \eqs{eq:dphi-dth}{eq:cphi-cth} are applicable again.
At $\phi \approx \phi_0$, the mass-squared of each orthogonal direction, defined as $m_i^2 \equiv d^2 V/ d i^2 \ (i=I, \psi)$, is found to be
\bea \label{eq:mI}
m_I^2 
%&\simeq& \frac{2 (pf)^2}{\l( qh \r)^2} \frac{V_0}{\phi^2}  \l[ 1-\l(1-\frac{q+1}{p} \r) \l( \frac{1}{f} - 1 \r) \r]
%\nonumber \\
&\simeq& m_{I,0}^2 \l( \frac{\phi}{\phi_0} \r)^{2p-2(q+1)}  \l[ 1-\l(1-\frac{1}{p} \r) \l( \frac{1}{f} - 1 \r) + \frac{q\kappa}{pf^2} \sin(h-\theta) \r]
\\ \label{eq:mpsi}
m_\psi^2 
%&\simeq& \frac{2 (pf)^2 V_0}{\phi^2} \l[ 1-\l(1-\frac{1}{p} \r) \l( \frac{1}{f} - 1 \r) \r] + (qh)^2 \frac{\Lambda^4}{\phi^2}
%\nonumber \\
&\simeq& m_{I,0}^2 (qh_0)^2 \l\{ \l( \frac{\phi}{\phi_0} \r)^{2(p-1)} \l[ 1-\l(1-\frac{1}{p} \r) \l( \frac{1}{f} - 1 \r) \r] + \kappa \frac{qh_0}{p} \l( \frac{\phi}{\phi_0} \r)^{2(q-1)} \r\}
\eea
where 
\bea \label{eq:mI0}
m_{I,0}^2 &\equiv& \frac{2 p^2}{\l( qh_0 \r)^2} \frac{V_0}{\phi_0^2} = 3 H_*^2 \times \frac{g(M, \phi_0)p}{p-(q+1)}
\\ \label{eq:mpsi0}
m_{\psi,0}^2 &\equiv& 2p^2 \l( 1 + \kappa \frac{qh_0}{p} \r) \frac{V_0}{\phi_0^2} = (qh_0)^2 \l( 1 + \kappa \frac{qh_0}{p} \r) m_{I,0}^2
\eea
In \eq{eq:mI} we have not applied \eq{eq:min-cond} for the term with $\sin(h-\theta)$ because the field configuration does not follow the trench unless it is trapped again on it.
During the oscillation phase which takes place mostly along the $\phi$ direction, the value of this term would vary. 

In \eq{eq:mI0}, $g(M, \phi_0)$ is a free parameter determined mainly by the set of $(M, \phi_0)$.
It can be either larger or smaller than unity, but lower-bounded as $g(M, \phi_0) > \l| \eta_* \r|$ for $p \geq 2(q+1)$ (see \eq{eq:eta}) which is the region we are interested in.
Hence, depending on $M$ and $\phi_0$, in the vicinity of $\phi \approx \phi_0$ we can have $g(M, \phi_0) > 1$ leading to $m_{I,0}^2 > 3 H_*^2$.
However, $m_I^2$ depends on $\phi$ and the slope along $I$ changes its sign across $\phi_0$.
As a result, the angular motion after inflation is commenced only when the oscillation amplitude along $\phi$ is significantly reduced so as to have $m_I^2 \gtrsim 3 H^2$.
In the vicinity of $\phi_0$, for the oscillation amplitude of $\phi$ denoted as $\delta \phi (\ll \phi_0)$, 
\beq
3 H^2 M_{\rm P}^2 \simeq \frac{1}{2} m_\phi^2 \l( \delta \phi \r)^2
\eeq
Hence,
\beq \label{eq:phi-oscL}
\frac{3H^2}{m_I^2} < 1 \Rightarrow \frac{\delta \phi}{\phi_0} < \frac{\delta \phi_{\rm osc}}{\phi_0} \equiv \frac{\sqrt{g(M, \phi_0)}}{p}
\eeq 
Therefore, the onset of the oscillation along $I$ is expected to happen as $\delta \phi$ is reduced to $\delta \phi_{\rm osc}$, and we find
\beq \label{eq:Hosc-gLow}
H_{\rm osc} = \sqrt{g(M, \phi_0)} H_*
\eeq 
which is valid only for $g(M, \phi_0) \ll 1$.
For $g(M, \phi_0) \gtrsim \mathcal{O}(1)$, we notice that $m_I^2$ changes its sign as the oscillation amplitude becomes smaller than 
\beq \label{eq:phi-oscH}
\frac{\delta \phi}{\phi_0} \simeq \frac{1}{2p} \Leftrightarrow f \simeq \frac{1}{2}
\eeq
and rapidly approaches to $m_{I,0}^2$.
We take this crossing point as the onset of the oscillation phase  of $I$ in this case, and the expansion rate around the epoch is found to be
\beq \label{eq:Hosc-gHigh}
H_{\rm osc} \approx \frac{1}{4 \sqrt{2}} H_*
\eeq 

A comment on the possibility of a second inflationary period  caused by re-trapping is in order.
From \eqss{eq:phi-retrap}{eq:phi-oscL}{eq:phi-oscH}, ${\rm Min}\l[ \sqrt{g(M, \phi_0)}, 1/2 \r] > \kappa$ and $m_I^2 > 3 H^2$ is expected around the epoch of re-trapping, i.e., the angular motion after inflation would take place before the field configuration is trapped in the trench.
Hence, a second stage of inflation would not take place.

Eventually, particles $I$ and $\psi$ would decay.
We express the decay rate of $i$-particle as \footnote{The precise form of $\Gamma_i$ depends on the couplings of each field to matter fields, but we do not specify those model-dependent couplings.}
\beq \label{eq:decay-rates}
\Gamma_i = \frac{\gamma_i}{8\pi} \frac{m_{i,0}^3}{\phi_0^2}
\eeq
where $\gamma_i (i=I,\psi)$ is a numerical constant taking allowed decay channels into account.
From \eqs{eq:mI}{eq:mpsi}, $m_{\psi,0} \gg m_{I,0}$ and generically we may expect $\psi$ decays earlier than $I$ as long as $\gamma_\psi \sim \gamma_I$.
Also, if $\Gamma_\psi \geq m_{I,0}$, the oscillation of $I$ field after inflation would take place in a universe dominated by radiation, otherwise it will happen in a universe dominated by $\psi$-particles.
In terms of our model parameters, the ratio of interest is given by 
\beq
\frac{\Gamma_\psi}{m_{I,0}} = \frac{3 p^2 \gamma_\psi}{4\pi} \frac{\sqrt{1+\kappa qh_0/p}}{qh_0} \l( \frac{H_*}{M_{\rm P}} \r)^2 \l( \frac{M_{\rm P}}{\phi_0} \r)^4
\eeq
If $\Gamma_\psi \gg \Gamma_I$, there is a possibility for $I$-particles to eventually dominate the universe around the epoch of its decay.  
In order to check this possibility, we compare the energy density of $I$-particles and that of the background radiation as follows.
When $I$ starts its oscillation, the oscillation amplitude is expected to be
\beq
I_{\rm osc} = \alpha \phi_0
\eeq
with $\alpha = \mathcal{O}(1) < \pi$.
As $\psi$ decays, the universe is dominated by radiation.
During this epoch, the energy density of $I$ before its decay is given by
\beq
\rho_I = \rho_I^{\rm osc} \l( \frac{a_{\rm osc}}{a_{I, \rm d}} \r)^3 
= \rho_I^{\rm osc} \l( \frac{H_{\psi, \rm d}}{H_{\rm osc}} \r)^2 \l( \frac{H}{H_{\psi, \rm d}} \r)^{3/2}
\eeq
where the energy density of $I$ at the onset of its oscillation is 
\beq
\rho_I^{\rm osc} = \frac{1}{2} m_{I, \rm osc}^2 I_{\rm osc}^2 \simeq \frac{3 \alpha^2}{2} H_{\rm osc}^2 \phi_0^2 
\eeq
and $H_{i, d} = (2/3) \Gamma_i$ with $i = (\psi, I)$ is the expansion rate at the epoch of $i$-particle decay.
Hence, $\rho_I$ becomes comparable to the background radiation density when $H=H_\times$ with
\beq
H_\times = \l[ \frac{\alpha^2}{2} \l( \frac{\phi_0}{M_{\rm P}} \r)^2 \r]^2 H_{\psi, \rm d}
\eeq
which gives
\bea
\frac{H_\times}{H_{I, \rm d}}
&=& \l[ \frac{\alpha^2}{2} \l( \frac{\phi_0}{M_{\rm P}} \r)^2 \r]^2 \frac{\Gamma_\psi}{\Gamma_I}
\nonumber \\
&=& \l[ \frac{\alpha^2 p \l[ p-(q+1) \r]}{g(M, \phi_0) (qh_0)^{1/2}} \l( \frac{\gamma_\psi}{\gamma_I} \r)^{1/2} \l( 1 + \frac{\kappa qh_0}{p} \r)^{3/4} \r]^2
\eea
Therefore, $I$ particles would be subdominant at the epoch of their decays only if
\beq
g(M, \phi_0) > \frac{\alpha^2 p \l[ p-(q+1) \r]}{(qh_0)^{1/2}} \l( \frac{\gamma_\psi}{\gamma_I} \r)^{1/2} \l( 1 + \frac{\kappa qh_0}{p} \r)^{3/4}
\eeq
Otherewise, an era of $I$-particle domination appears, and entropy release due to the lat-time decay of $I$ takes place.
This causes a dilution of pre-existing particle densities relative to entropy density.
In the sudden decay approximation, the dilution factor $\Delta$ is approximately given by 
\beq \label{eq:dilution}
\Delta \approx \frac{T_\times}{T_I} = \l( \frac{g_*(T_I)}{g_*(T_\times)} \r)^{1/4} \l[ \frac{\alpha^2}{2} \l( \frac{\phi_0}{M_{\rm P}} \r)^2 \r] \l( \frac{\Gamma_\psi}{\Gamma_I} \r)^{1/2}
\eeq
where $T_\times$ and $T_I$ are respectively temperatures when $H=H_\times$ and $H_{I, \rm d}$.

\section{Charge asymmetry after inflation}
\label{sec:asymmetry}
In this section, we consider the possibility of generating a $B$- or $L$-asymmetry from the dynamics or decay of the $I$-field.
From now on, we assume that the fields $\phi$ and $\theta$ in the potential \eq{eq:V} are components of a complex field such as $\Phi \equiv \phi e^{i\theta} / \sqrt{2}$ which can carry a charge denoted as $Q$.

\subsection{Spontaneous baryogenesis}

A slow motion of $I$ after inflation might be considered for spontaneous baryogenesis by introducing a derivative interaction of $\theta$ to baryonic/leptonic current in Lagrangian such as
\beq
\mathcal{L} \supset \lambda \l( \partial_\mu \theta \r) j^\mu \simeq \frac{\lambda}{\phi_0} \l( \partial_\mu I \r) j^\mu
\eeq
where $\lambda$ is a dimensionless coupling constant, and $j^\mu$ is a baryonic/leptonic current.
In the current scenario we are considering, a slow evolution of $\theta$ after inflation takes place around the epoch of $H \sim H_{\rm osc}$ which is given by either \eq{eq:Hosc-gLow} or (\ref{eq:Hosc-gHigh}).
Soon after this epoch, the dynamics of $I$ turns into rapid oscillations.
If $B$ or $L$ violating processes which might have been already in equilibrium are decoupled at the very epoch, it might be possible for spontaneous baryogenesis to work.
%But those processes should be decoupled at a right epoch, i.e., $H\sim H_{\rm osc}$, otherwise $I$ falls into the oscillating phase before the decoupling.
However, if $\Gamma_\psi < H_{\rm osc}$, the universe around the epoch of $I$'s oscillation would be only partially reheated by the partial decay of $\psi$ particles.
This means that the background temperature between the end of inflation and onset of $I$'s oscillation is expected to be
\beq
T_{\rm osc} \lesssim T \lesssim T_e 
\eeq
where 
\bea
T_e &\sim& \l( \Gamma_\psi H_* M_{\rm P}^2 \r)^{1/4}
\\
T_{\rm osc} &\sim& \l( \Gamma_\psi H_{\rm osc} M_{\rm P}^2 \r)^{1/4} \sim 
\l\{
\begin{array}{ll}
g(M, \phi_0)^{1/4} T_e &\textrm{if} \ g(M, \phi_0) \ll 1
\\
2^{-5/4} T_e &\textrm{if} \ g(M, \phi_0) \gtrsim 1
\end{array}
\r.
\eea
That is, $T_{\rm osc}$ and $T_e$ differ only by a factor of a few since $g(M, \phi_0) \gtrsim \l| \eta_* \r| \sim \mathcal{O}(10^{-2})$.
Also, as $\Gamma_\psi$ becomes closer to or even larger than $H_{\rm osc}$, the epoch between $T_e$ and $T_{\rm osc}$ becomes narrower.
Hence, spontaneous baryogenesis is unlikely to occur in our scenario.

\subsection{Chage asymmetry from the decay of $I$ particles}

As another possibility of generating a charge asymmetry, which was already discussed in the original paper of spontaneous baryogenesis (Ref. \cite{Cohen:1987vi}), we consider the decay of $I$ at its oscillation phase, assuming $B$ or $L$-violating processes were decoupled already before the onset of $I$'s oscillations.
Here we do not specify $B$ or $L$-violating operators, but simply assume the branching fraction of relevant channels to be close to unity. 

%In this case, the angular motion corresponds to charge asymmetry with a given sign determined by the direction of the motion.
%
%The evolution equations are given by
%\bea
%3H^2 M_{\rm P}^2 &=& \rho_\psi + \rho_I + \rho_{\rm r}
%\\
%\dot{\rho_\psi} + 3H \rho_\psi &=& - \Gamma_\psi \rho_\psi
%\\
%\dot{\rho_{\rm r}} + 4H \rho_{\rm r} &=& \Gamma_\psi \rho_\psi
%\\
%\ddot{I} + \l( 3H + \Gamma_I \r) \dot{I} + m_I^2 I &=& 0
%\eea
%where $\rho_{\rm r}$ is the energy density of radiation produced mainly in the process of the decay of $\psi$-particles whose energy density is $\rho_\psi$, and $\rho_I$ are the energy densityof $I$.
During the era of $\psi$-particle domination with the sudden decay approximation of $\psi$, approximately the solution of $I$'s EOM can be taken to be \footnote{The error in the estimation of late-time charge asymmetry is less than a factor about 2, and does not affect our argument.}
\beq
I = I_{\rm osc} a^{-3/2} e^{- \frac{\Gamma_I}{2}t} \cos \l[m_{I,0} \l( t - t_{\rm osc} \r) \r]
\eeq
and the charge density associated with $I$ is given by
\bea
n_Q 
&=& -i Q \l( \Phi^* \dot{\Phi} - \dot{\Phi}^* \Phi \r) = Q \dot{\theta} \phi^2 \simeq Q \dot{I} \phi_0
\nonumber \\
&=& - \frac{Q}{2} \l\{ 3H+\Gamma_I + 2m_{I,0} \tan \l[ m_{I,0} \l( t - t_{\rm osc} \r) \r] \r\} I \phi_0
\eea 
where we used $\phi \simeq \phi_0$ and $\theta \simeq I/\phi_0$ in the vicinity of the true vacuum.
Then, the charge asymmetry from the decay of $I$ is obtained as \footnote{This is essentially of the same form as one discussed in Ref.~\cite{Cohen:1987vi} modulo the factor $\mathcal{F}(x_{\rm osc}, x)$.}
\beq \label{eq:nB}
n_B 
= \frac{1}{a^3} \int_{t_{\rm osc}}^{t} dt \Gamma_I \l( a^3 n_Q \r)
= - \frac{Q}{a^3} \Gamma_I I_{\rm osc} \phi_0 \mathcal{F}(x_{\rm osc}, x)
\eeq
where $x \equiv m_{I,0} t$ and 
\bea \label{eq:integral-F}
\mathcal{F}(x_{\rm osc}, x) 
&\equiv& \int_{x_{\rm osc}}^{x} dx \ e^{ - \frac{\Gamma_I}{2 m_{I,0}} x} \l( \frac{x}{x_{\rm osc}} \r)^{\frac{1}{1+\omega}} \l\{ \frac{1}{(1+\omega) x} + \frac{\Gamma_I}{2 m_{I,0}} + \tan \l( x- x_{\rm osc} \r) \r\} \cos \l( x - x_{\rm osc} \r) 
\nonumber \\
&\stackrel{x \to \infty}{\longrightarrow}& e^{-\frac{\Gamma_I x_{\rm osc}}{2 m_{I,0}}} + \frac{e^{-i x_{\rm osc}} E_{\frac{w}{1+w}}\l( \frac{(\Gamma_I/m_{I,0} - 2i) x_{\rm osc}}{2} \r) + e^{i x_{\rm osc}} E_{\frac{w}{1+w}}\l( \frac{(\Gamma_I/m_{I,0} + 2i) x_{\rm osc}}{2} \r)}{1+w}
\nonumber \\
&\stackrel{\Gamma_I \lll m_{I,0}}{\longrightarrow}& 1 + \frac{e^{-i x_{\rm osc}} E_{\frac{w}{1+w}}\l( -i x_{\rm osc} \r) + e^{i x_{\rm osc}} E_{\frac{w}{1+w}}\l( i x_{\rm osc} \r)}{1+w}
\nonumber \\
&\simeq& \l\{
\begin{array}{ll}
1.5 &: w=1/3
\\
1 &: w=0
\end{array}
\r.
\eea
with 
\beq
x_{\rm osc} = \frac{2}{3(1+\omega)} \frac{m_{I,0}}{H_{\rm osc}} = \mathcal{O}(1)
\eeq
and $\omega$ being the equation of state of the universe.
The approximation in the second line of \eq{eq:integral-F} is valid for $x \gg m_{I,0} /\Gamma_I$ (i,e,, $t \gg 1/\Gamma_I$), and $E_\nu(z)$ is \textit{Exponential Integral E} function with a complex argument.
%, which is defined as
%\beq
%E_\nu(z) \equiv \int_1^{\infty} dt \frac{e^{-zt}}{t^\nu} \ ; \ \textrm{Re}[z] \geq 0
%\eeq
Thus, for $t \gg 1/\Gamma_I$ one finds
\beq \label{eq:nB}
n_B \simeq - \frac{Q}{a^3} \Gamma_I I_{\rm osc} \phi_0 \mathcal{F}(x_{\rm osc}, \infty)
%= -Q \l( \frac{H}{H_{\rm osc}} \r)^2 \Gamma_I I_{\rm osc} \phi_0 \mathcal{F}(x_{\rm osc}, \infty)
\eeq
Meanwhile, if the energy density of $I$-particles is subdominant around the epoch of $I$-particle decay, the entropy density after inflation evolves as 
\bea
s &=& s_\psi \l( \frac{a_\psi}{a} \r)^3
\\
s_\psi &=& \beta \l( \frac{H_\psi}{H_{\rm osc}} \r)^{3/2} \l( H_{\rm osc} M_{\rm P} \r)^{3/2}
\eea
where $s_\psi$ is the entropy density around the epoch of $\psi$-particle decays,
\beq
\beta \equiv \frac{2 \pi^2 g_{*S}(T_\psi)}{45} \l( \frac{\pi^2 g_*(T_\psi)}{90} \r)^{-3/4} \simeq 8.66 \times \l( \frac{g_*}{200} \r)^{1/4}
\eeq
with $T_\psi$ being the background temperature at $\psi$-particle decay, and we set $g_{*S}(T) = g_*(T_\psi)$, ignoring their temperature-dependence.
%---------------------
\begin{figure}[t]
\begin{center}
\includegraphics[width=0.45\textwidth]{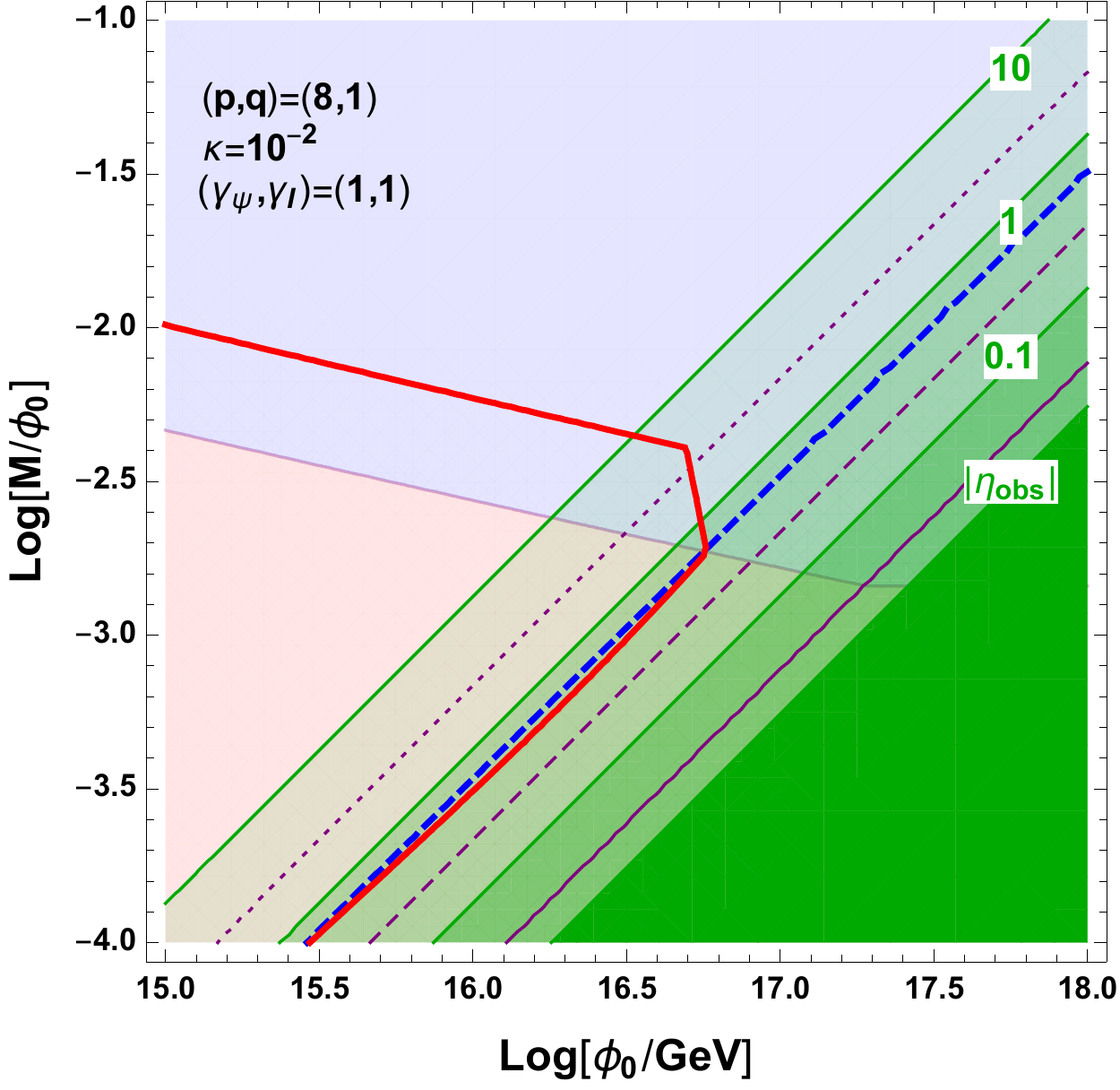}
\includegraphics[width=0.45\textwidth]{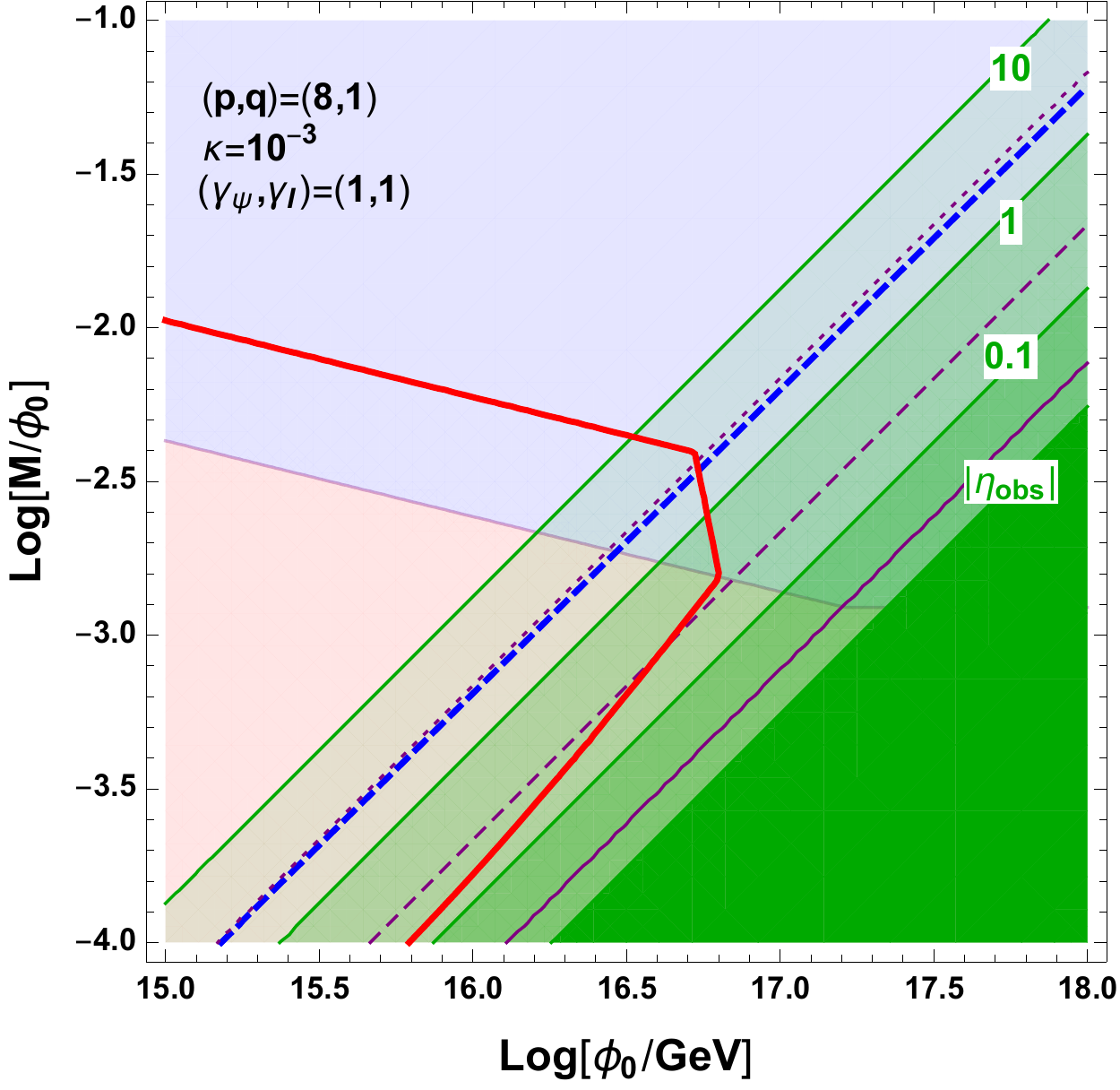}
\caption{Parameter space for baryon number asymmetry (red-line) as a function of $M$ and $\phi_0$.
Color scheme other than blue and red lines is the same as Fig.~\ref{fig:large-p-parameters}.
The light blue and pinky regions divided by a thin light purple line are regions of $H_\psi > H_{\rm osc}$ and $H_\psi < H_{\rm osc}$, respectively.
The dashed blue line indicates parameters matching observed values of inflationary observables $n_s$ and $P_R$ for a given $\kappa$ which  satisfies \eq{eq:kappa}.
The solid red line is for $(n_B/s)/(Q \alpha \mathcal{F})= \l( n_B/s \r)^{\rm obs} = 8.7 \times 10^{-11}$ \cite{pdg-bbn}.
The line below the lower sharp breaking point is for $H_\psi > H_{\rm osc}$.
The line above the upper sharp breaking point is for $H_\times < H_{I, \rm d}$. 
Taking a smaller or larger $\gamma_I \alpha$ shifts the red line left or right side for a given $\kappa$.
}
\label{fig:nbs}
\end{center}
\end{figure}
%---------------------- 
Hence, at late time 
\beq \label{eq:nBs-latetime}
\l| \frac{n_B/s}{Q} \r| 
\simeq \mathcal{F}(x_{\rm osc}, \infty) \frac{\gamma_I \alpha}{8 \pi \beta} \l( \frac{m_{I,0}}{M_{\rm P}} \r)^{3/2} \l( \frac{m_{I,0}}{H_{\rm osc}} \r)^{3/2} \l( \frac{H_\psi}{H_{\rm osc}} \r)^{1/2}
\eeq
For $p=2(q+1)$, using \eqss{eq:eta}{eq:mI0}{eq:Hosc-gLow}, one can re-express \eq{eq:nBs-latetime} as
\bea \label{eq:nBs-L}
\l| \frac{n_B/s}{Q} \r| 
&\simeq& \mathcal{F}(x_{\rm osc}, \infty) \frac{9 \gamma_I \alpha}{2 \pi \beta} |\eta_*|^{3/2} \l( \frac{H_*}{M_{\rm P}} \r)^{3/2} \l( \frac{H_\psi}{H_{\rm osc}} \r)^{1/2}
\nonumber \\
&\simeq& 1.0 \times 10^{-10} \mathcal{F}(x_{\rm osc}, \infty) \gamma_I \alpha \l( \frac{g_*}{200} \r)^{-1/4} \l( \frac{H_*}{10^{14} \GeV} \r)^{3/2} \l( \frac{H_\psi}{H_{\rm osc}} \r)^{1/2}
\eea
For $p > 2(q+1)$, using \eqs{eq:mI0}{eq:Hosc-gHigh}, one can re-express \eq{eq:nBs-latetime} as
\bea \label{eq:nBs-H}
\l| \frac{n_B/s}{Q} \r| 
&\simeq& 2^{15/4} \mathcal{F}(x_{\rm osc}, \infty) \frac{\gamma_I \alpha}{8 \pi \beta} \l[ \frac{3p g(M, \phi_0)}{p-(q+1)} \r]^{3/2} \l( \frac{H_*}{M_{\rm P}} \r)^{3/2} \l( \frac{H_\psi}{H_{\rm osc}} \r)^{1/2}
\nonumber \\
&\simeq& 2.7 \times 10^{-9} \mathcal{F}(x_{\rm osc}, \infty) \gamma_I \alpha \l( \frac{g_*}{200} \r)^{-1/4} \l[ \frac{p g(M, \phi_0)}{p-(q+1)} \r]^{3/2} \l( \frac{H_*}{10^{13} \GeV} \r)^{3/2}  \l( \frac{H_\psi}{H_{\rm osc}} \r)^{1/2}
\eea
If $I$-particle dominates the universe eventually before its decay, \eqs{eq:nBs-L}{eq:nBs-H} should be divided by the dilution factor given in \eq{eq:dilution}.

If $p=2(q+1)$, $H_* \ll H_*^{\rm bnd}$ for $\kappa \ll 1$ as shown in Fig.~\ref{fig:low-p-parameters}. 
Hence, in this case, even with $H_*$ close to its upper-bound and $H_\psi > H_{\rm osc}$, an additional  requirement, namely  $\gamma_I \alpha \gtrsim \mathcal{O}(1)$ is needed in order to obtain $n_B/s \sim 10^{-10}$.  

On the other hand, if $p \gg 2(q+1)$, for a much lower $H_*$ one can find parameter space for a right amount of baryon/lepton number asymmetry as shown in Fig.~\ref{fig:nbs} where we took $(p,q)=(8,1)$ as a benchmark set.
In the figure, the crossing point of dashed blue line and solid red line fixes parameters matching observations of $n_s, P_R$, and $n_B/s$ simultaneously.
For $(\gamma_\psi,\gamma_I)=(1,1)$ taken in the figure, if $\kappa \lesssim 10^{-2}$, the matching point appears in $\psi$-particle domination era.
A smaller $\gamma_\psi$ extends $\psi$-domination era, pushing the border of light blue and pinky regions to left-bottom side of each panel.
A smaller $\gamma_I$ results in a smaller baryon asymmetry, pushing the upper and lower part of the red line to bottom and left side, respectively. 
%For a given $\kappa$, the crossing point spans some ranges of $M$ and $\phi_0$, depending on $\gamma_I \alpha$. 
As a kind of guide line, if $\gamma_I \lesssim \gamma_\psi \lesssim 1$, in order to obtain a right amount of asymmetry it is required to have $p \gtrsim 8$ with $q=1$ for a reasonable choice of $\kappa$, such as $\kappa \sim \mathcal{O}(10^{-3\pm1})$.
A larger $q$ requires even larger $p$ and much smaller $\kappa$.
Such a $p$-dependence of the asymmetry can be understood as follows.
For given set of parameters other than $p$, a larger $p$ causes a smaller $\eta$ (see \eq{eq:eta}).
Hence, in order to match observations, i.e., $\eta = \eta_*$, one should make $(M_{\rm P}/\phi_0) (M/\phi_0)^q$ larger by either increasing $M$ or decreasing $\phi_0$.
This increases $m_{I,0}$ (see \eq{eq:mI}) and $\Gamma_I$.
As a result, a larger asymmetry can be obtained, which is clear from the second line of \eq{eq:nB}.

%
%
%\subsection{A naive estimation}
%\beq
%n_Q = \dot{\theta} \phi_0^2 = \dot{I} \phi_0 \sim m_I I_{\rm osc} \phi_0
%\eeq
%\beq
%\Gamma_I = \frac{\gamma_I}{8 \pi} \frac{m_I^3}{\phi_0}
%\eeq
%Then, from the decay of $I$, 
%\beq
%n_B \sim \frac{\Gamma_I}{H_{\rm osc}} n_Q \sim \frac{\Gamma_I}{m_I} m_I I_{\rm osc} \phi_0 = \frac{\alpha \gamma_I}{8 \pi}  m_I^3
%\eeq
%Hence, in RD,
%\beq
%\frac{n_B}{s} \sim \frac{n_B}{(m_I M_{\rm P})^{3/2}} \sim \frac{\alpha \gamma_I}{8 \pi} \l( \frac{m_I}{M_{\rm P}} \r)^{3/2}
%\eeq
%For 
%\beq
%m_I \lesssim m_I^{\rm max} \sim 10^{13} \GeV
%\eeq
%a right amount of matter-antimatter asymmetry can be obtained only if
%\beq
%\alpha \gamma_I \sim \gamma_I \sim 8 \pi \l( \frac{n_B}{s} \r)_{\rm obs} \l( \frac{m_I}{M_{\rm P}} \r)^{-3/2} \sim 0.3 \l( \frac{10^{13} \GeV}{m_I} \r)^{3/2}
%\eeq
%Note that it is likely to have $\gamma_I \lesssim \mathcal{O}(100)$.
%Hence, $m_I \gtrsim \mathcal{O}(10^{12}) \GeV$ is required.
%However, question is if such a large $m_I$ can be consistently obtained while the universe is dominated by radiation.
%

\section{Conclusions}
\label{sec:con}

In this paper, we investigated the possibility of generating an asymmetry of a global charge (e.g., baryon/lepton number) through either the  dynamics or the decays of the inflaton in the scenario of \textit{spiral inflation}.

In spiral inflation, the inflaton can be regarded dominantly as the angular degree of a complex scalar field.
Thanks to the presence of a small angle-dependent modulating potential on top of a hilltop-like potential, it gets through a spiraling-out motion from the hilltop.
Although inflation ends via a waterfall-like sudden change of field dynamics, angular motion of field configuration reappears after inflation again due to the presence of the angle-dependent modulating potential.
When it carries a non-zero global charge, the angular motion of a complex scalar field corresponds to a charge asymmetry (or particle-antiparticle asymmetry) associated with the field.
Hence, the angular motion of inflaton after inflation implies an asymmetry associated with inflaton number density.
The angular momentum of inflaton in the vicinity of the true vacuum of the potential does not posses a definite sign, but periodically changes its sign.
As a result, the asymmetry generated by  inflaton changes its sign periodically.
However, as discussed in the original paper of spontaneous baryogenesis (Ref.~\cite{Cohen:1987vi}), if the inflaton decays to other particles, transferring its asymmetry (say `transfer mechanism'), there can be a well-defined (net) asymmetry in the daughter particles, thanks to the expansion of the universe.

Paying attention to the transfer mechanism, we found that in spiral inflation the decays of the inflaton can produce the right amount of baryon number asymmetry while obtaining inflationary observables consistent with observations.
%The late-time baryon number asymmetry in our scenario depends on 
%\begin{itemize}
%\item[] $\Lambda^4/V_0$ - the ratio of heights of two potentials ($\Lambda^4$: modulation, $V_0$: hilltop)
%\item[] $M/\phi_0$ - the ratio of scales appearing in two potentials
%\item[] $\phi_0$ - the scale of symmetry-breaking in hilltop potential
%\end{itemize}
%in addition to the exponents $p$ and $q$ determining curvatures of background hilltop potential and modulating one, and (as expected) model-dependent decay rate (or branching fraction) of inflaton to baryonic/leptonic particles through operators (maybe) violating  baryon/lepton number.  
In contrast to the naive expectation that it would be difficult to obtain the right amount of baryon number asymmetry, it is found that even in the presence of matter-domination era a sufficient amount of baryon number asymmetry can be obtained as long as the matter-domination era right after inflation is terminated rather soon.
Definitely this is a model-dependent statement, since (as expected) the late-time baryon number asymmetry in our scenario depends on model-dependent decay rate(s) (or branching fraction(s)) of the inflaton to baryonic/leptonic particles through operators (maybe) violating  baryon/lepton number.

In a part of the parameter space, the expansion rate during inflation is required to be close to the current upper-bound, and hence it would be easily probed in the near-future experiments, for example CMB-S4 \cite{cmb-s4}, PIXIE \cite{Kogut:2011xw}, and LiteBIRD \cite{LiteBIRD}.

\acknowledgments
GB acknowledges support from the MEC and FEDER (EC) Grants SEV-2014-0398, FIS2015-72245-EXP, and FPA2014-54459  and the Generalitat Valenciana under grant PROMETEOII/2017/033. She also acknowledges partial support from the European Union FP7 ITN INVISIBLES MSCA PITN-GA-2011-289442 and InvisiblesPlus (RISE) H2020-MSCA-RISE-2015-690575. 
WIP was supported by Research Base Construction Fund Support Program funded by Chonbuk National University in 2018, and by Basic Science Research Program through the National Research Foundation of Korea (NRF) funded by the Ministry of Education (No. 2017R1D1A1B06035959).

%\paragraph{Note added.} This is also a good position for notes added after the paper has been written.

%\appendix



\begin{thebibliography}{99}



% Please avoid comments such as "For a review'', "For some examples",
% "and references therein" or move them in the text. In general,
% please leave only references in the bibliography and move all
% accessory text in footnotes.

% Also, please have only one work for each \bibitem.

\bibitem{pdg-bbn}
M. Tanabashi et al. (Particle Data Group), Phys. Rev. D 98, 030001 (2018)

\bibitem{Aghanim:2018eyx} 
  N.~Aghanim {\it et al.} [Planck Collaboration],
  %``Planck 2018 results. VI. Cosmological parameters,''
  arXiv:1807.06209 [astro-ph.CO].


%%  Inflation 
%\cite{Guth:1980zm}
\bibitem{Guth:1980zm} 
  A.~H.~Guth,
  %``The Inflationary Universe: A Possible Solution to the Horizon and Flatness Problems,''
  Phys.\ Rev.\ D {\bf 23}, 347 (1981).
  %%CITATION = PHRVA,D23,347;%%
  
%\cite{Sato:1980yn}
\bibitem{Sato:1980yn} 
  K.~Sato,
  %``First Order Phase Transition of a Vacuum and Expansion of the Universe,''
  Mon.\ Not.\ Roy.\ Astron.\ Soc.\  {\bf 195}, 467 (1981).
  %%CITATION = MNRAA,195,467;%%
  
%\cite{Starobinsky:1980te}
\bibitem{Starobinsky:1980te} 
  A.~A.~Starobinsky,
  %``A New Type of Isotropic Cosmological Models Without Singularity,''
  Phys.\ Lett.\ B {\bf 91}, 99 (1980).
  %%CITATION = PHLTA,B91,99;%%


\bibitem{Sakharov:1967dj} 
  A.~D.~Sakharov,
  %``Violation of CP Invariance, C asymmetry, and baryon asymmetry of the universe,''
  Pisma Zh.\ Eksp.\ Teor.\ Fiz.\  {\bf 5}, 32 (1967)
  [JETP Lett.\  {\bf 5}, 24 (1967)]
  [Sov.\ Phys.\ Usp.\  {\bf 34}, no. 5, 392 (1991)]
  [Usp.\ Fiz.\ Nauk {\bf 161}, no. 5, 61 (1991)].


\bibitem{Cohen:1987vi} 
  A.~G.~Cohen and D.~B.~Kaplan,
  %``Thermodynamic Generation of the Baryon Asymmetry,''
  Phys.\ Lett.\ B {\bf 199}, 251 (1987).
  doi:10.1016/0370-2693(87)91369-4

\bibitem{Cohen:1988kt} 
  A.~G.~Cohen and D.~B.~Kaplan,
  %``Spontaneous Baryogenesis,''
  Nucl.\ Phys.\ B {\bf 308}, 913 (1988).
  doi:10.1016/0550-3213(88)90134-4
  %%CITATION = doi:10.1016/0550-3213(88)90134-4;%%


\bibitem{DeSimone:2016ofp} 
  A.~De Simone and T.~Kobayashi,
  %``Cosmological Aspects of Spontaneous Baryogenesis,''
  JCAP {\bf 1608}, no. 08, 052 (2016)
  doi:10.1088/1475-7516/2016/08/052
  [arXiv:1605.00670 [hep-ph]].


\bibitem{Anber:2015yca} 
  M.~M.~Anber and E.~Sabancilar,
  %``Hypermagnetic Fields and Baryon Asymmetry from Pseudoscalar Inflation,''
  Phys.\ Rev.\ D {\bf 92}, no. 10, 101501 (2015)
  doi:10.1103/PhysRevD.92.101501
  [arXiv:1507.00744 [hep-th]].

\bibitem{Cado:2016kdp} 
  Y.~Cado and E.~Sabancilar,
  %``Asymmetric Dark Matter and Baryogenesis from Pseudoscalar Inflation,''
  JCAP {\bf 1704}, no. 04, 047 (2017)
  doi:10.1088/1475-7516/2017/04/047
  [arXiv:1611.02293 [hep-ph]].

  
\bibitem{Takahashi:2015ula} 
  F.~Takahashi and M.~Yamada,
  %``Spontaneous Baryogenesis from Asymmetric Inflaton,''
  Phys.\ Lett.\ B {\bf 756}, 216 (2016)
  doi:10.1016/j.physletb.2016.03.020
  [arXiv:1510.07822 [hep-ph]].


\bibitem{Barenboim:2014vea} 
  G.~Barenboim and W.~I.~Park,
  %``Spiral Inflation,''
  Phys.\ Lett.\ B {\bf 741}, 252 (2015)
  doi:10.1016/j.physletb.2014.12.042
  [arXiv:1412.2724 [hep-ph]].

\bibitem{Barenboim:2015zka} 
  G.~Barenboim and W.~I.~Park,
  %``Spiral Inflation with Coleman-Weinberg Potential,''
  Phys.\ Rev.\ D {\bf 91}, no. 6, 063511 (2015)
  doi:10.1103/PhysRevD.91.063511
  [arXiv:1501.00484 [hep-ph]].






\bibitem{cmb-s4}
https://cmb-s4.org/

\bibitem{Kogut:2011xw} 
  A.~Kogut {\it et al.},
  %``The Primordial Inflation Explorer (PIXIE): A Nulling Polarimeter for Cosmic Microwave Background Observations,''
  JCAP {\bf 1107}, 025 (2011)
  doi:10.1088/1475-7516/2011/07/025
  [arXiv:1105.2044 [astro-ph.CO]].

\bibitem{LiteBIRD}
http://litebird.jp/eng/


\end{thebibliography}
\end{document}